\documentclass[preprint,draft,showpacs,aps,fleqn]{revtex4}%
\usepackage{graphicx}
\usepackage{bm}
\usepackage{dcolumn}

\begin{document}
\def\be{\begin{eqnarray}}
\def\ee{\end{eqnarray}}
\newcommand{\no}{\nonumber}
\def\mpcomm#1{\nextline\strut\kern-6em{\tt MP COMMENT => \ #1}\nextline}
\def\nextline{\hfill\break}

\preprint{TPJU-4/2004~~~BNL-NT-04/10}

\title{Chiral doublers  of heavy-light baryons}

\author{Maciej A. Nowak$^{a)}$, Micha\l{} Prasza\l{}owicz$^{a,b)}$,\\
Mariusz Sadzikowski$^{a)}$ and Joanna Wasiluk$^{a)}$\\
~~~}

\affiliation{$^{a)}$M. Smoluchowski Institute of Physics,
Jagiellonian
University,\\
30-059 Krak\'{o}w, Reymonta 4, Poland. \\
$^{b)}$Nuclear Theory Group, Brookhaven National Laboratory,\\
Upton, NY 11973-5000, USA. \\~~~~}

\begin{abstract}
We discuss  the consequences of  the chiral doubling
scenario for baryons  built of heavy {\em and}
light quarks. In particular, we use the soliton description for
baryons, demonstrating why each heavy-light baryon should be
accompanied by the opposite parity partner. Our argumentation
holds both for ordinary baryons and for exotic heavy pentaquarks
which are required by the symmetries of QCD to appear in parity
doublets, separated by the mass  shift of the chiral origin.
Interpreting the recently observed by BaBaR,
CLEO and Belle charmed mesons with
assignment $(0^+,1^+)$ as the chiral partners of known $D$ and
$D^*$ mesons, allows us to estimate the parameters of the mesonic
effective lagrangian, and in consequence, estimate the masses of
ground states and excited states of both parities. In particular,
we interpret the state recently reported by the H1 experiment at
HERA as a chiral partner $\tilde{\Theta}_c^0(3099)$
of {\em yet} undiscovered ground state pentaquark
$\Theta_c^0(2700)$.
\end{abstract}


 \maketitle

Recently, experimental physics of hadrons with open charm
has provided several spectacular discoveries:\\
-- First, BaBar~\cite{BB} has announced new, narrow meson
 $D^{*}_{sJ} (2317)^+$,
decaying into $D_s^+$ and $\pi^0$. This observation was then confirmed by
CLEO~\cite{CLEO}, which also noticed another narrow state,
 $D_{sJ}(2463)^+$, decaying
into $D_s^*$ and $\pi^0$. Both states were confirmed
by Belle~\cite{BELLE}, and finally,
the CLEO observation was also confirmed by BaBar~\cite{BABAR2}.\\
-- Second, Belle has not only measured the narrow excited states
$D_1,D_2$ with foreseen  quantum numbers $(1^+,2^+)$, but provided
also first evidence for two new, broad states $D_0^*$ $(2308\pm
17\pm 15\pm 28)$ and $D_1^{'}$ $(2427\pm 26 \pm 20 \pm 17)$
\cite{Bellenonstrange}. Both
of them  are approximately $400$ MeV  above the usual $D_0,D^{*}$
states and seem to have opposite
to them parity.\\
-- Third, Selex  has provided preliminary data for
   doubly charmed baryons~\cite{SELEX}. On top of known since 2002
   $ccd$ state $(3520)$, four other cascade $j=1/2$ states are visible,
in particular the pair of opposite parity $ccu$ states
separated by the mass gap of the order $337$MeV.\\
-- Fourth, H1 experiment at DESY has announced~\cite{H1DESY} a
signature for
   charmed {\it pentaquark} $\bar{c}udud$ at mass 3099 MeV,
   {\em i.e.} approximately
   400 MeV higher than the expected estimates known in the
literature~\cite{OPM,JW,KarLip,MAWU}.

   The above states  and in particular the decay patterns
of all this particles are a challenge for   standard estimations
based on quark potential models  and triggered a flurry of
activity among the theorists.

An appealing possibility is that the presence of the above states
is the consequence of so-called chiral doublers scenario,
theoretically anticipated~\cite{us92,bh93} already in 1992 and
1993. In brief,  the scenario, based on simultaneous constraints
of the spontaneous breakdown of chiral symmetry for light quarks
and of heavy quark spin symmetry (Isgur-Wise symmetry)~\cite{IW}
for heavy quarks, leads to parity  duplication of all heavy-light
hadrons, with mass shift fixed by the pattern of the spontaneous
breakdown of the chiral symmetry. Very recently, in  light of
Babar, CLEO and Belle discoveries, chiral doublers scenario was
reminded~\cite{newchir}, pointing  that the $D_s (2317)$ is a
$0^+$ chiral partner of the usual $(0^-)$ $D_s$ state and the
$D_s^*(2463)$ is a $1^+$ partner of the known vector $D^*_s$.
Similar arguments were suggested for new $(0^+,1^+)$ states
observed by Belle~\cite{Bellenonstrange} and similar kind of
doubling is also expected for excited non-strange heavy-light
mesons.

In this note, we discuss the extension of the chiral doublers
scenario for all baryons, including the exotic states
(pentaquark). To avoid any new parameters, we simply view baryons
as solitons of the effective mesonic lagrangian including {\em
both} chiral copies of heavy-light mesons, 
a point addressed  already in~\cite{us92}. We are working in large
$N_c$ limit, which justifies the soliton picture, and large heavy
quark mass limit, where we exploit the Isgur-Wise symmetry. This
approach could be viewed as a starting point for including $1/m_h$
corrections from the finite mass of the heavy quark, explicit
breaking of chiral symmetry, etc.

The description of baryons as solitons of the mesonic lagrangians
has a long history. Original Skyrme~\cite{SKYRME} idea was
elaborated by Witten~\cite{WITTEN}, and Adkins, Nappi and
Witten~\cite{WAN} for $SU(2)_{flavor}$ with enormous success and
hundreds of followers. The extension to SU(3) was more tricky.
Simple embedding of Skyrme ansatz and proper inclusion of the
Wess-Zumino-Witten-Novikov term led  to the appearance of octet,
decuplet and antidecuplet baryons~\cite{SKYRMIONSU3}. On one side,
this approach, including the explicit mass of the strange quark
seemed to be less successful phenomenologically then the $SU(2)$
version~\cite{MPcom}. On the other side, it is one of very few
{\em dynamical} models, which predicted the presence of strange
pentaquark~\cite{PENTAS}, a state widely discussed nowadays and
observed by several experiments~\cite{EXPERPENTAS}.

The mixed success of the first order perturbation theory in $m_s$
in the Skyrme model~\cite{MPcom} led in two directions. One
consisting in enriching the splitting hamiltonian by terms
subleading in $N_c$~\cite{CQM} gave very good description of
hyperon mass spectra and produced successful prediction of the
strange pentaquark mass. The other one was based on the simple
observation that the mass of the strange quark is of the same
order as the inverse moment of inertia of the soliton. This fact
tempted Callan and Klebanov ~\cite{CK} to consider an alternative
scheme for $SU(3)$. They looked at the binding of the kaon in the
field of the $SU(2)$ soliton, and then collectively quantized the
bound state as a whole. Although this approach was
phenomenologically successful, its extension to strange
pentaquarks revealed some fundamental difficulties
\cite{KlebTheta}.

The bound state approach was expected to work even better for
baryons including heavy quark, {\em i.e.} the charm one for
example~\cite{heavy}. In
~\cite{JMW} it was pointed out that such an approach does not
respect the Isgur-Wise symmetry -- for infinitely heavy quarks the
soliton should bound the degenerate in the IW limit pair of a
pseudoscalar and a vector, i.e. $D$ and $D^*$. Charmed hyperons
emerge therefore as bound states of $D$ and $D^*$ in the presence
of the SU(2) Skyrme background. The related
approach~\cite{NRZBOOK} used the Born-Oppenheimer approximation.
First the pseudoscalar-vector heavy meson pair was bound in the
background of the static soliton, generating the $O(N_c^0)$
binding. Vibrational modes were the ``fast degrees'' of the
freedom. The adiabatical rotation of the bound system by
quantization of collective coordinates of the $SU(2)$ skyrmions
corresponds then to ``slow degrees'' of freedom. It is well known,
that in this case the rotation is  not the free one. Fast degrees
of freedom in Born-Oppenheimer approximation generate the
effective ``gauge'' potential, of a Berry phase~\cite{wilczekzee}
type. In the case of degenerate pesudoscalar and vector mesons (IW
limit) the phases coming from  $D$ meson and $D^*$ meson are
equal, but opposite. Their cancellation corresponds to the
realization of the Isgur-Wise symmetry at the baryonic level,
therefore degeneration of spin 1/2 and 3/2 multiplets.

In the following, we choose this philosophy, but contrary to other
approaches known in literature~\cite{JMW,OPM,MAWU}, we consider
the full heavy-light effective lagrangian with both chiral
copies~\cite{us92,bh93}. A related  approach was considered 
in~\cite{HARADA}, however the effects of chiral shift were 
not included. To avoid unnecessary repetitions we
rewrite this lagrangian using the conventions applied in solitonic
calculations in the $D$ meson sector~\cite{OPM}. The full
lagrangian reads now
 \be L=L_H + L_G+L_{GH}
 \label{total}
 \ee
 where
  \be
 L_H&=&-i {\rm
  Tr}(\bar{H}v^{\mu}D_{\mu}H) +   g_H{\rm Tr}
   H\gamma^{\mu}\gamma_5 A_{\mu} \bar{H} \nonumber \\
    &&+m_H(\Sigma)\,{\rm Tr}\bar{H}\,H
\label{usualcopy}
 \ee
 is the usual~\cite{wise} lagrangian (modulo the last $O(m_h^0)$ mass term,
depending on constituent mass of the light quark~\cite{us92,bh93})
for the standard $(0^-,1^-)$ multiplet
 \be
H= \frac{1+v\!\!\!/}{2}( \gamma_5D-\gamma^{\mu}D^{*}_{\mu} )
 \ee
and
 \be
 L_G& = &-i {\rm
  Tr}(\bar{G}v^{\mu}D_{\mu}G) +   g_G{\rm Tr}
   G\gamma^{\mu}\gamma_5 A_{\mu} \bar{G}\nonumber \\
    &&+ m_G(\Sigma)\,{\rm Tr}\bar{G}\,G
\label{notsousualcopy}
 \ee
 is the chiral doubler lagrangian for $(0^+,1^+)$ chiral partner
 \be
G= \frac{1+v\!\!\!/}{2}(
\tilde{D}-\gamma^{\mu}\gamma_5\tilde{D}^{*}_{\mu})\,\,.
 \ee
 Chiral partners  communicate with  each other via light axial
currents
 \be
L_{HG}=
g_{GH}{\rm Tr}(\gamma_5\bar{G}H \gamma^{\mu} A_{\mu}) + (h.c.)
\label{int}
 \ee
with no vector mixing because of the parity.
The axial $A_{\mu}$ reads
 \be
A_{\mu} =\frac{i}{2}(\xi^{\dagger}\partial_{\mu} \xi
- \xi \partial_{\mu}\xi^{\dagger})
 \ee
 where $\xi^2=U=\exp (i \vec{\pi} \cdot \vec{\tau})$ and $v_{\mu}$
is the four-velocity of the heavy quark. In our case, we take the
pion field as the Skyrme hedgehog ansatz $\pi_i=F(r)n_i$.

 The key difference in the chiral copy is the opposite sign of
the constituent mass contribution in (\ref{notsousualcopy})
$m_G(\Sigma)\approx -\Sigma$, with respect to the similar term for
the $H$ multiplet, $m_H(\Sigma)\approx \Sigma$, where $\Sigma$
denotes one loop heavy meson self-energy~\cite{us92,bh93,newchir}.
The sign flip follows from the $\gamma_5$ difference in the
definition of the fields $H$ and $G$. In other words:  it is
sensitive to the parity content of the heavy-light field since
$H\rlap/{v}=-H$ and $G\rlap/{v}=+G$. The result is a split between
the heavy-light mesons of opposite chirality.

Standard approach~\cite{JMW,OPM,MAWU}
ignores (``integrates'') the heavier chiral copy $G$, and the heavy
hyperon spectrum comes only from $L_H$ part of the lagrangian
leading to~\cite{OPM}
 \be
M=M_{sol} +m_D -3/2 g_H F'(0) +a/I_1
 \ee
 where $M_{sol}$ is the $O(N_c)$ classical mass of the Skyrmion,
$m_D=(3M_{D^*}+M_D)/4$ is the averaged mass of heavy-light mesons,
$g_H$ is the axial coupling constant responsible for the $D^*$
decays into a D and a pion, and the inverse of moment of inertia
of the Skyrmion $1/I_1$ provides the splitting between the various
isospin states. We follow here the conventions of~\cite{OPM}. Since
for isosinglet $a=3/8$ and for isotriplet $a=11/8$, one
immediately recovers the remarkable formula~\cite{JMW}
 \be
M(\Sigma_h)-M(\Lambda_h)=\frac{1}{I_1}=\frac{2}{3}(M(\Delta)-M(N))
 \ee
where the r.h.s. comes from the $SU(2)$ Skyrme model.

The pentaquark spectrum comes~\cite{OPM} from
 replacing the meson by antimeson
in the field of soliton with baryon number one, and for the
isosinglet pentaquark mass formula reads:
 \be
M_5=M_{sol} +m_D -1/2 g_H F'(0) +3/(8I_1)
 \ee
so pentaquark  in this model is three times less bound that the
heavy hyperon. Since numerically~\cite{OPM} $M_{sol}=866$~MeV,
averaged $m_D=1973$ MeV, binding strenght $g_H F'(0)=419$ MeV,
$I_1^{-1}=195$ MeV, the estimate for the $\Theta_c=\bar{c}udud$
pentaquark mass is 2702 MeV, in agrement with recent estimates of
the correlated quark model~\cite{JW} and SU(3) soliton
calculations~\cite{MAWU}.

The spectrum of pentaquarks~\cite{OPM} is strongly degenerate in
mass. However, there is no mixing between these states because
they differ in parity of the state, the parity of the light
degrees of freedom and/or isospin. The mixing is suppressed by the
powers of the heavy quark mass and by the number of colors.

Let us consider now the full lagrangian~(\ref{total}). First we
observe, that due to the properties of the heavy spin symmetry,
one can trade $\gamma^{\mu} A_{\mu}$ into $v^{\mu}A_{\mu}$ in
(\ref{int}). This implies, that in the rest frame {\em static}
Skyrmion background decouples the $G$ and $H$ lagrangians. Similar
observation holds for the version of binding in the scenario of
ref.~\cite{JMW},
 where the coupling vanishes due to the $r\delta(r)$ term from the
wave function of the infinitely heavy charmed meson. This
decoupling allows immediately to write down the mass formula for
opposite chirality partner of the isoscalar baryon and for
opposite chirality partner of the isoscalar pentaquark (denoted by
tilde)
 \be
\tilde{M}  &=& M_{sol} +m_{\tilde{D}} -3/2 g_G F'(0) +3/(8I_1) \nonumber \\
\tilde{M}_5&=& M_{sol} +m_{\tilde{D}} -1/2 g_G F'(0) +3/(8I_1)
 \ee
 It is of primary importance that, despite the additional
$\gamma_5$ in the definition of the $G$ field
(\ref{notsousualcopy}), both hamiltonians have the same functional form 
of lowest
eigenvalue: $M_5$ for $H$ and $\tilde{M}_5$ for $G$. Hence both
chiral partners emerge as $H$ and $G$  bound states in the SU(2)
solitonic background. The mass difference comes in the first
approximation solely from the difference of the coupling constants
$g_G-g_H$  and meson mass difference $m_{\tilde{D}}-m_D$ where
$m_{\tilde{D}}=(3M_{\tilde{D*}}+M_{\tilde{D}})/4$ is the averaged
over heavy-spin mass of the $(1^+,0^+)$ mesons. Constant $g_G$ is
the axial coupling constant in the opposite parity channel,
responsible for pionic decays of the $1^+$ axial states into $0^+$
scalars. Using recent Belle data~\cite{BELLE}, {\em i.e.} $0^+$
candidate  $D_0^*$ $(2308\pm 17\pm 15\pm 28)$ and $1^+$ candidate
$D_1^{'}$ $(2427\pm 26 \pm 20 \pm 17)$, we get
$M_{\tilde{D}}=2397$ MeV, unfortunately with still large errors.

In the case of strange $D_s$, the impressive evidence  for such
states comes from BaBar~\cite{BB}, Cleo~\cite{CLEO} and
Belle~\cite{BELLE} data.

Let us combine now the above formulae. Fist, we notice, that the
 mass splitting between the usual baryons of opposite
parity leads to
 \be \Delta_B= \Delta_M  +3/2F'(0)g_H \delta g
  \ee where
$\Delta_M=M_{\tilde{D}}-M_D$ is the mass shift between the
opposite parity heavy-light mesons and $\delta_g=1-g_G/g_H$
measures the difference between the axial couplings for both
copies. Similar reasoning leads to the formula for the chiral
splitting between the opposite parity pentaquarks:
 \be
\Delta_P= \Delta_M  +1/2F'(0)g_H \delta g\; .
 \ee Combining both
formulae we get
 \be \Delta_P=\frac{\Delta_B +2 \Delta_M}{3}\; .
 \ee

Let us turn now towards the data. Comparing the mass shift between
the lowest $\Lambda_c$ states of opposite parities,
$\Lambda_c(1/2^+, ~2285)$ and $\Lambda_c(1/2^-,~2593)$ we arrive
at $\Delta_B$ = 310 MeV. Similarly, $\Xi_c(1/2^+,~2470)$ and
$\Xi_c(1/2^-,~2790)$ give $\Delta_B$ = 320 MeV. Comparing the
shift of the opposite parity heavy charmed mesons from very recent
Belle~\cite{Bellenonstrange} data we arrive at $\Delta_M=425$~MeV
unfortunately with still large errors. These two numbers allow us
to estimate $\Delta_P=350$~MeV~$\pm 60$~MeV, {\em i.e.} we get the
mass of the chiral partner of the pentaquark as high as $3052 \pm
60$~ MeV. We note that the argument proposed here is based on the
leading approximation in large $N_c$ and large $m_h$ limit, and is
intended to demonstrate the order of magnitude for chiral
splitting for heavy pentaquarks.

One is therefore tempted to interpret the recent H1
state~\cite{H1DESY} as a chiral partner $\tilde{\Theta}_c$ of the
yet undiscovered isosinglet pentaquark $\Theta_c$ of opposite
parity and $M_5\approx 2700$~MeV. Similar reasoning applies to
other isospin channels, stranged charmed pentaquarks and to
extensions for b quarks. Despite Babar and Cleo data yield with
the impressive accuracy the chiral mesonic shift to be equal to
350 MeV, no charmed strange baryon data for both parities do exist
by now, so one cannot make similar estimation for strange charmed
pentaquarks.

Unfortunately present accuracy of the soliton models does not
allow to estimate splittings between pentaquarks of various
parities and spins~\cite{OPM}. Assuming that spin 3/2 pentaquarks
will be shifted in mass by $1/m_h$ corrections, there are still
two $1/2^-$ and one $1/2^+$ degenerate isosinglet states of the
$(D,D^*)$ mesons bound in the soliton background, and similarly
three 1/2 states of opposite parities in the
$(\tilde{D},\tilde{D}^*)$ sector.

Understanding the narrow width of the new state reported by H1
remains a challenge. Our scenario offers, however, a qualitative
explanation. Let us first observe that the natural channel for the
decay of this state into a nucleon and {\em chiral partners} of
the standard $D (D^*)$ mesons is kinematically blocked.
De-excitation of $\tilde{\Theta}_c$ into $\Theta_c$ and a pion is
isospin forbidden and to $\eta^0$ kinematically blocked. Three body
decay into  $\Theta_c\;2\pi$ has very small phase space. Therefore
the only way the decay process may proceed, is a chiral
fluctuation of a bound $\tilde{D}$ into $D$ by virtual interaction
with a pion from the nucleon cloud. That requires, however,
spacial rearrangement, since the $D$ meson must be in a partial
wave of opposite parity with respect to the partial wave of
$\tilde{D}$ . Hence the overlap of the $\tilde{D}$-soliton bound
state wave function with the one of the  $D$-soliton is expected
to be small.

In this note, we pointed out that the surprisingly heavy mass of
the new charmed pentaquark state may be naturally interpreted in
the chiral doubler scenario, forcing each heavy-light hadron to
have the opposite parity partner. This pattern seems to be
confirmed by now for strange charmed mesons, and is very likely
for the recently observed charmed non-strange mesons. In the
baryonic sector the universal shift of approximately $310\pm
10$~MeV seems to separate heavy-light-light and heavy-heavy-light
conventional states with opposite parity. If heavy pentaquarks
exist, similar pattern of chiral doubling forces them to appear in
opposite parity pairs.

\begin{acknowledgments}
MP thanks  S.H. Kahana and D.E. Kahana for useful conversations.
 This work was partially supported
by the Polish State Committee for Scientific Research (KBN) grant
2P03B 096 22 (MAN, JW), 2P03B 093 22 (MS) and 2 P03B 043 24 (MP). This manuscript
has been authored under Contract No. DE-AC02-98CH10886 with the U.
S. Department of Energy.
\end{acknowledgments}


{
}

\end{document}